\documentclass[final]{cvpr}
\usepackage{times}
\usepackage{epsfig}
\usepackage{graphicx}
\usepackage{amsmath}
\usepackage{amssymb}
\usepackage{xcolor}
\usepackage{multirow}
\usepackage{multicol}
\usepackage{subfigure, enumitem}
\usepackage{threeparttable, booktabs}
\usepackage[export]{adjustbox}

\usepackage[pagebackref=true,breaklinks=true,colorlinks,bookmarks=false]{hyperref}

\def\cvprPaperID{****} 

\begin{document}

\title{NTIRE 2021 Challenge on Quality Enhancement of Compressed Video:\\ Dataset and Study}

\author{Ren Yang\\Computer Vision Laboratory\\ETH Z\"urich, Switzerland\\{\tt\small ren.yang@vision.ee.ethz.ch}
\and
Radu Timofte\\Computer Vision Laboratory\\ETH Z\"urich, Switzerland\\{\tt\small radu.timofte@vision.ee.ethz.ch}
}

\maketitle

\begin{abstract}
This paper introduces a novel dataset for video enhancement and studies the state-of-the-art methods of the NTIRE 2021 challenge on quality enhancement of compressed video. The challenge is the first NTIRE challenge in this direction, with three competitions, hundreds of participants and tens of proposed solutions. Our newly collected Large-scale Diverse Video (LDV) dataset is employed in the challenge. In our study, we analyze the proposed methods of the challenge and several methods in previous works on the proposed LDV dataset. We find that the NTIRE 2021 challenge advances the state-of-the-art of quality enhancement on compressed video. The proposed LDV dataset is publicly available at the homepage of the challenge: \url{https://github.com/RenYang-home/NTIRE21_VEnh}
\end{abstract}


\section{Introduction}\label{intro}

During the recent years, there is increasing popularity of video streaming over the Internet~\cite{Cisco} and the demands on high-quality and high-resolution videos are also rapidly increasing. Due to the limited bandwidth of the Internet, video compression~\cite{wiegand2003overview, sullivan2012overview} plays an important role to significantly reduce the bit-rate and facilitate transmitting a large number of high-quality and high-resolution videos. However, video compression unavoidably leads to compression artifacts, thus resulting in the loss of both fidelity and perceptual quality and the degradation of Quality of Experience (QoE). Therefore, it is necessary to study on enhancing the quality of compressed video, which aims at improving the compression quality at the decoder side. Due to the rate-distortion trade-off in data compression, enhancing compressed video is equivalent to reducing the bit-rate at the same quality, and hence it also can be seen as a way to improve the efficiency of video compression.

\begin{figure}[!t]
\includegraphics[width=1.07\linewidth, right]{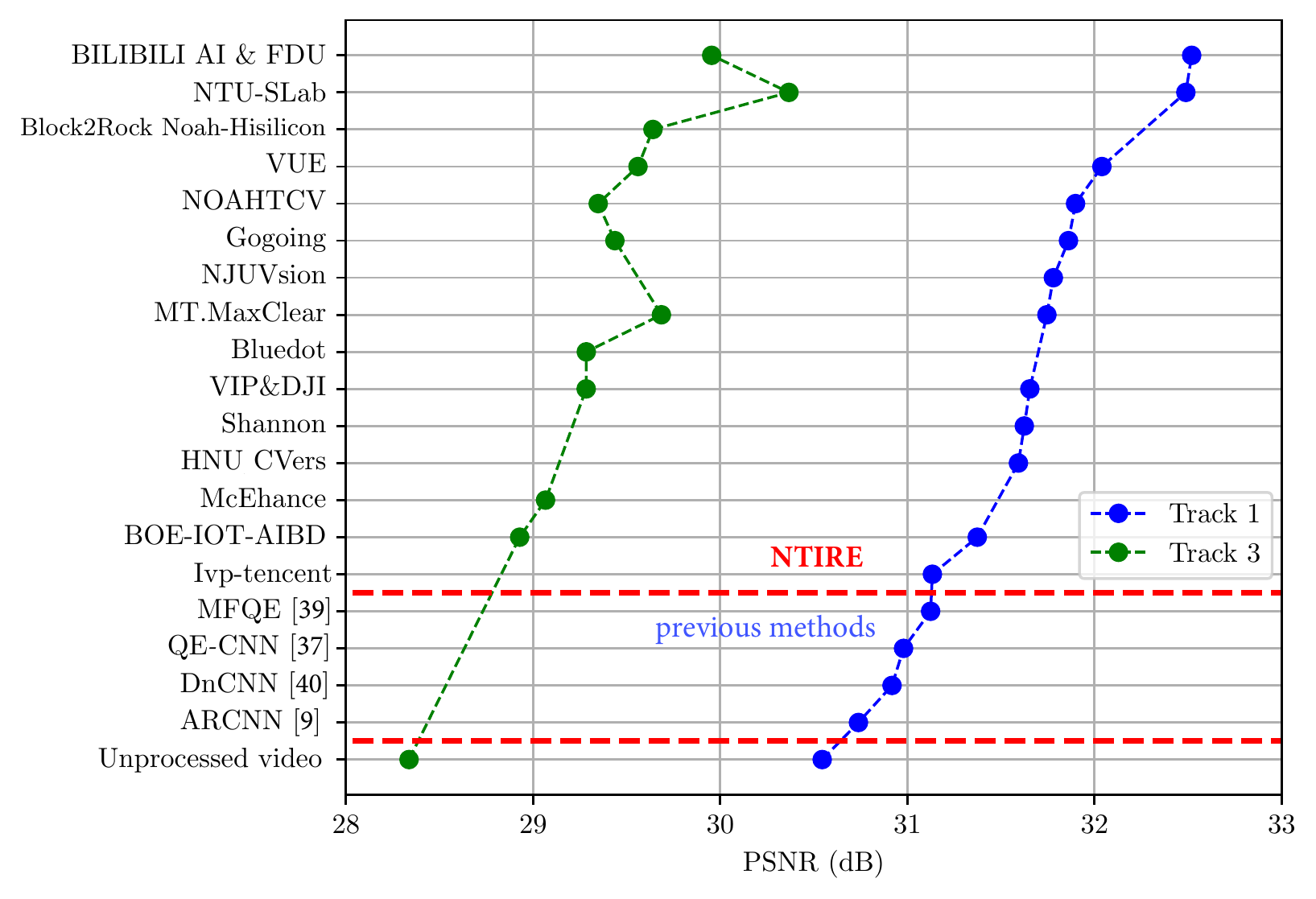}
\caption{The performance on the fidelity tracks.}
\label{fig:1}
\end{figure}

In the past decade, a great number of works~\cite{yang2017decoder, yang2018enhancing, wang2017novel, lu2018deep, yang2018multi, xu2021multi, yang2019quality, guan2019mfqe, Xu_2019_ICCV, deng2020spatio, yang2020learning, huo2021recurrent, wang2020multi} have been proposed in this direction. In these works, \cite{yang2017decoder, yang2018enhancing, wang2017novel} propose enhancing compression quality based on a single frame, and \cite{lu2018deep, yang2018multi, yang2019quality, guan2019mfqe, Xu_2019_ICCV, deng2020spatio, yang2020learning, huo2021recurrent, wang2020multi} are multi-frame quality enhancement methods. Besides, Wang~\etal~\cite{wang2020multi} propose enhancing compressed video towards perceptual quality, which is evaluated by the Mean Opinion Score (MOS). Other works~\cite{yang2017decoder, yang2018enhancing, wang2017novel, lu2018deep, yang2018multi, yang2019quality, guan2019mfqe, Xu_2019_ICCV, deng2020spatio, yang2020learning, huo2021recurrent} focus on improving the Peak Signal-to-Noise Ratio (PSNR) for better fidelity performance.

\begin{figure*}[!t]
\centering
\includegraphics[width=1\linewidth]{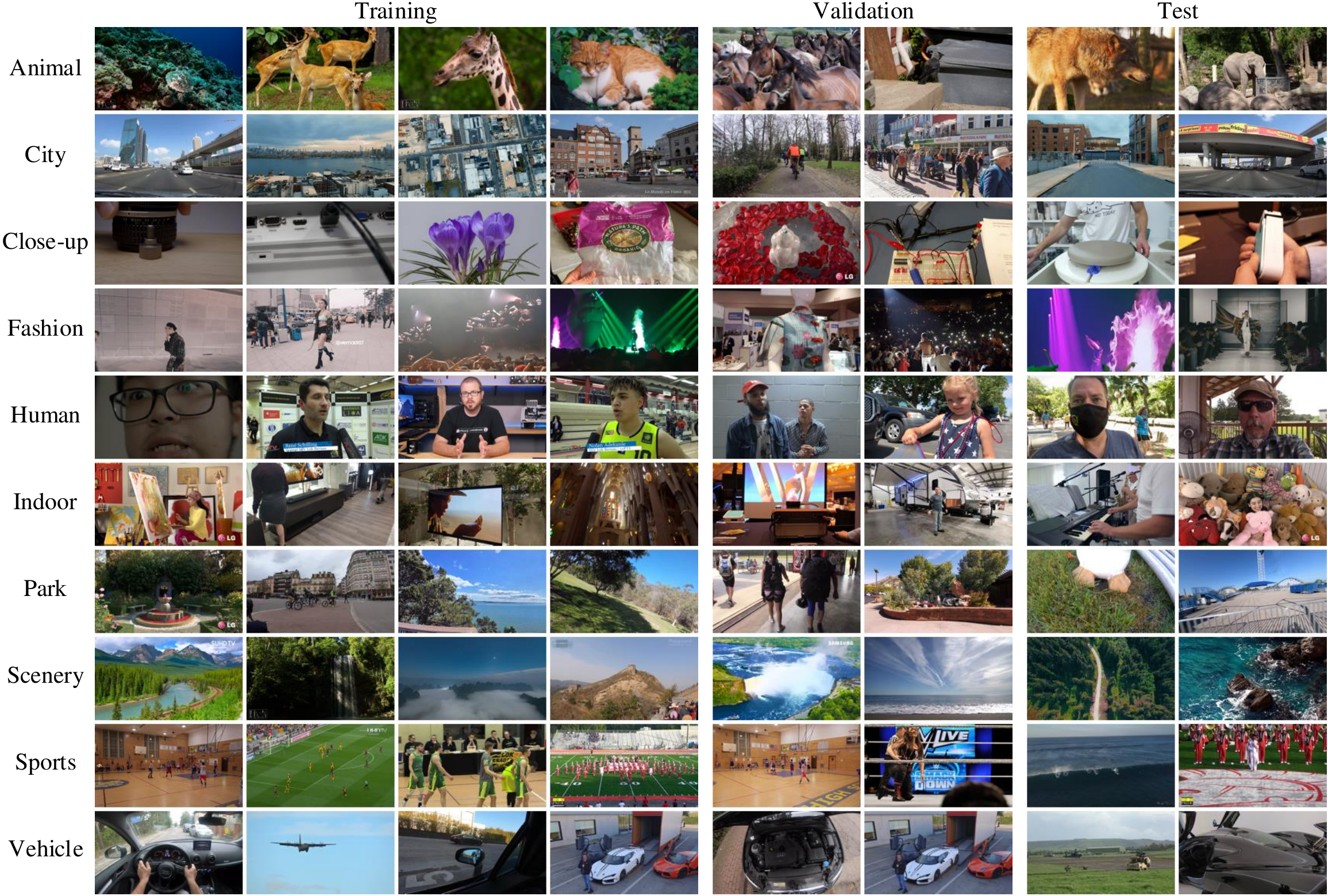}
\caption{Example videos in the proposed LDV dataset, which contains 10 categories of scenes. The left four columns show a part of videos used for training in the NTRIE challenge. The two columns in the middle are the videos for validation. The right two columns are the test videos, in which the left column is the test set for Tracks 1 and 2, and the most right column is the test set for Track 3.}
\label{fig:dataset}
\vspace{-1em}
\end{figure*}

Thanks to the rapid development of deep learning~\cite{lecun2015deep}, all aforementioned methods are deep-learning-based and data-driven. In single-frame enhancement literature, \cite{wang2017novel} uses the image database BSDS500~\cite{arbelaez2010contour} as the training set, without any video. \cite{yang2017decoder} trains the model on a small video dataset including 26 video sequences, and then \cite{yang2018enhancing} enlarges the training set to 81 videos. Among the existing multi-frame works, the model of \cite{lu2018deep} is trained on the Vimeo-90K dataset~\cite{xue2019video}, in which each clip only contains 7 frames, and thus it is insufficient for the research on enhancing long video sequences, especially not applicable for recurrent frameworks. Then, the Vid-70 dataset, which includes 70 video sequences, is used as the training set in \cite{yang2018multi, yang2019quality, Xu_2019_ICCV, huo2021recurrent}. Meanwhile, \cite{guan2019mfqe} and \cite{deng2020spatio} collected 142 and 106 uncompressed videos for training, respectively. \cite{wang2020multi} uses the same training set as \cite{guan2019mfqe}. In conclusion, all above works train the models on less than 150 video sequences, except Vimeo-90K which only has very short clips. Therefore, establishing a large scale training database with high diversity is essential to promote the future research on video enhancement. Besides, the commonly used test sets in existing literature are the JCT-VC dataset~\cite{bossen2013common} (18 videos), the test set of Vid-70~\cite{yang2019quality} (10 videos) and Vimeo-90K (only 7 frames in each clip). Standardizing a larger and more diverse test set is also meaningful for the deeper studies and fairer comparisons for the proposed methods.

\begin{figure*}[!t]
\centering
\includegraphics[width=1\linewidth]{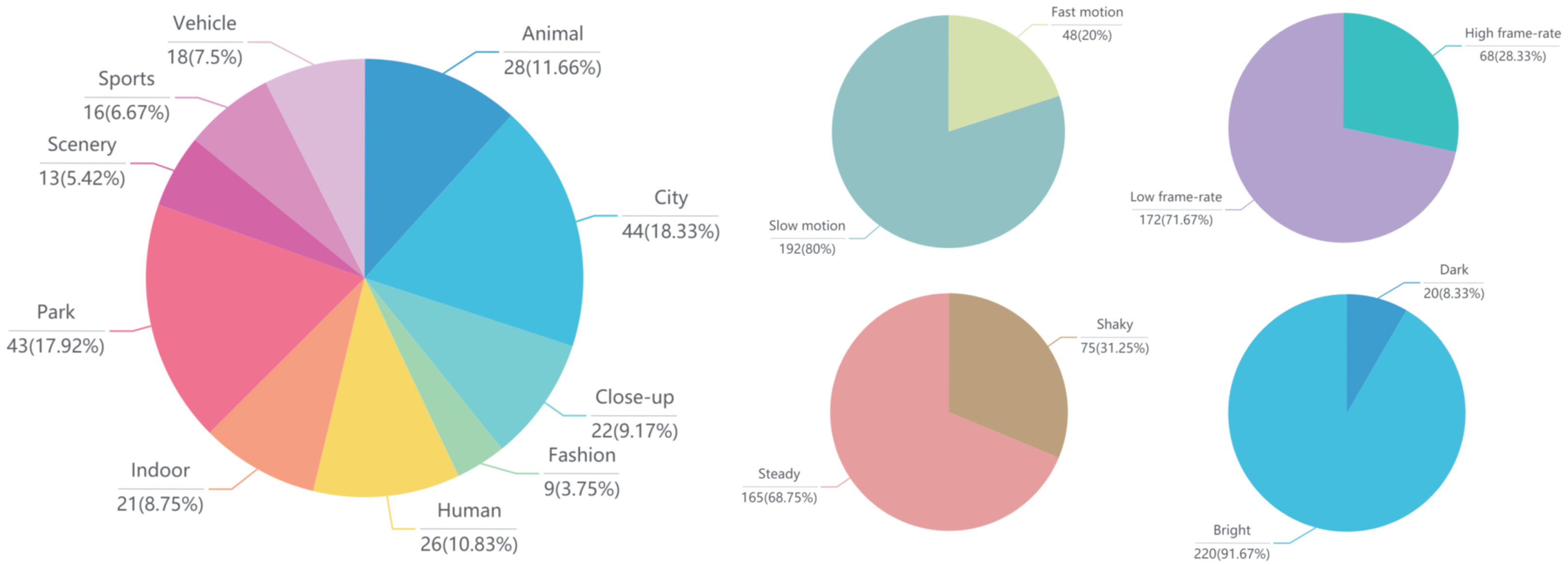}
\caption{The diversity of the proposed LDV dataset.}
\label{fig:div}
\end{figure*}

In this work, we collect a novel Large-scale Diverse Video (LDV) dataset, which contains 240 high quality videos with the resolution of $960\times536$\footnote{A few videos are with the resolution of $960\times504$ or $960\times496$.}. The proposed LDV dataset includes diverse categories of contents, various kinds of motion and different frame-rates. Moreover, we organized the online challenge at NTIRE 2021\footnote{\url{https://data.vision.ee.ethz.ch/cvl/ntire21/}} for enhancing the quality of compressed video using the LDV dataset. The challenge has three tracks. In Tracks 1 and 2, videos are compressed by HEVC~\cite{sullivan2012overview} at a fixed QP, while in Track 3, we enable rate-control to compress videos by HEVC at a fixed bit-rate. Besides, Tracks 1 and 3 aim at improving the fidelity (PSNR) of compressed video, and Track 2 targets at enhancing the perceptual quality, which is evaluated by the MOS value.

Moreover, we also study the newly proposed LDV dataset via the performance of the solutions in the NTIRE 2021 video enhancement challenge and the popular methods from existing works. Figure~\ref{fig:1} shows the PSNR performance on the fidelity tracks. It can be seen from Figure~\ref{fig:1} that all NTIRE methods effectively improve the PSNR of compressed video, and the top methods obviously outperforms the the existing methods. The PSNR improvement of NTIRE methods on Track 1 ranges from 0.59 dB to 1.98 dB, and on Track 3 ranges from 0.59 dB to 2.03 dB. We also report the results in various quality metrics in Section 5 for analyses. 

The remainder of the paper is as follows. Section II introduces the proposed LDV dataset, and Section III describes the video enhancement challenge in NTIRE 2021. The methods proposed in the challenge and some classical existing methods are overviewed in Section IV. Section V reports the challenge results and the analyses on the LDV dataset.

\section{The proposed LDV dataset}

We propose the LDV dataset with 240 high quality videos with diverse content, different kinds of motion and large range of frame-rates. The LDV dataset is intended to complement the existing video enhancement datasets to enlarge the scale and increase the diversity to establish a more solid benchmark. Some example videos in the proposed LDV dataset are shown in Figure~\ref{fig:dataset}.

\textbf{Collecting.} The videos in LDV are collected from YouTube\footnote{\url{https://www.youtube.com/}}. To ensure the high quality, we only collect the videos with 4K resolution, and without obvious compression artifacts. All source videos used for our LDV dataset have the licence of \emph{Creative Commons Attribution licence (reuse allowed)}\footnote{\url{https://support.google.com/youtube/answer/2797468}}. Note that the LDV dataset is for academic and research proposes.

\textbf{Diversity. } We mainly consider the diversity of videos in our LDV dataset from three aspects: category of scenes, motion and frame-rate. As Figure~\ref{fig:dataset} illustrates, our LDV dataset contains 10 categories of scenes, \ie, \textit{animal}, \textit{city}, \textit{close-up}, \textit{fashion}, \textit{human}, \textit{indoor}, \textit{park}, \textit{scenery}, \textit{sports} and \textit{vehicle}. In the 240 videos of LDV, 48 videos are with fast motion. The frame-rates of the videos in LDV range from 24 to 60, in which 172 videos are with low frame-rates ($\leq30$) and 68 videos are with high frame-rates ($\geq50$). Additionally, the camera is slightly shaky (\eg, captured by handheld camera) in 75 videos of LDV. The shakiness results in irregular local movement of pixels, which is a common phenomenon especially in the videos of social media (cameras hold by hands). Besides, 20 videos of LDV are in dark environments, \eg, at night or in the rooms with insufficient light. 

\textbf{Downscaling. } To remove the compression artifacts of the source videos, we downscale the videos by the factor of $4$ using the Lanczos filter~\cite{turkowski1990filters}. Then, the width and height of each video are cropped to the multiples of 8, due to the requirement of the HEVC test model (HM). We follow the standard datasets, \eg, JCT-VC~\cite{bossen2013common}, to convert videos to the format of YUV 4:2:0.

\textbf{Partition. } In the challenge of NTIRE 2021, we divide the LDV dataset into three parts for training (200 videos), validation (20 videos) and test sets (20 videos), respectively. The 20 test videos are further split into two sets with 10 videos each. The two test sets are used for the track of fixed QP (Tracks 1 and 2) and the track of fixed bit-rate (Track 3), respectively. The validation set contains the videos from the 10 categories of scenes with two videos in each category. Each test set has one video from each category. Besides, 9 in the 20 validation videos and 4 of the 10 videos in each test set are with high frame-rates. There are five fast-motion videos in the validation set, and there are three and two fast-motion videos in the test sets for fixed QP and fixed bit-rate tracks, respectively.  

\section{NTIRE 2021 video enhancement challenge}

The video enhancement challenge of NTIRE 2021~\cite{yang2021ntire} has three tracks. Tracks 1 and 2 target at enhancing the quality of video compressed by HM 16.20 at fixed QP ($=37$), and Track 3 is for the enhancement of video compressed by x265 at fixed bit-rate ($=200$ kbps). Besides, Tracks 1 and 3 aim at improving the fidelity of compressed video, and Track 3 aims at enhancing compressed video towards perceptual quality. The websites of the three tracks are at: 

Track 1: \url{https://competitions.codalab.org/competitions/28033}

Track 2: \url{https://competitions.codalab.org/competitions/28034}

Track 3: \url{https://competitions.codalab.org/competitions/28035}

\subsection{Video Quality Assessment (VQA)}\label{VQA}

\textbf{Fidelity. } In the fidelity tracks, we evaluate the quality of enhanced video by PSNR and the Multi-Scale Structural SIMilarity index (MS-SSIM)~\cite{wang2003multiscale}. In our challenge, PSNR and MS-SSIM are calculated in the RGB domain on each frame, and then they are averaged among all frames of each video. The rank of proposed methods in the fidelity tracks depends on PSNR, and we additionally report MS-SSIM for analyses.  

\textbf{User study. } In the track of perceptual quality, the proposed methods are ranked by the MOS values~\cite{mos}. We set the range of MOS values as 0 to 100. There are 15 subjects participating in the MOS experiment. We show the uncompressed videos to the subjects as the standard of $s=100$. However, the subjects are asked to rate the enhanced videos according to their visual quality, instead of the similarity to the uncompressed videos. Then, defining  $s_{max}$ and $s_{min}$ as the highest and the lowest score of each subject, respectively, the scores ($s$) of each subject are linearly normalized to
\begin{equation}\label{norm}
    s' = 100 \cdot \frac{s-s_{min}}{s_{max} - s_{min}}.
\end{equation}
To ensure the rating consistency of each subject, we insert five repeated videos to check the rating consistency, and the scores of the four least consistent subjects (average error on repeated videos $\geq 20$) are omitted. Therefore, the final MOS values are obtained as the average of 11 subjects.

\textbf{Perceptual metrics. } In addition to the user study, we further evaluate the methods for perceptual quality by various perceptual metrics, including the Learned Perceptual Image Patch Similarity (LPIPS)~\cite{zhang2018unreasonable}, Fr\'echet Inception Distance (FID)~\cite{heusel2017gans}, Kernel Inception Distance (KID)~\cite{binkowski2018demystifying} and Video Multimethod Assessment Fusion (VMAF)~\cite{VMAF}. LPIPS~\cite{zhang2018unreasonable} measures the distance in the feature space of DNN. FID~\cite{heusel2017gans} and KID~\cite{binkowski2018demystifying} calculates the similarity between the distributions of the groundtruth and distorted frames. VMAF~\cite{VMAF} is developed by Netflix for assessing the perceptual quality of video. These metrics have been validated to be effective for evaluating perceptual quality. 

\subsection{Tracks 1 and 2: Fixed QP}

In Tracks 1 and 2, videos are compressed by the official HEVC test model HM 16.20\footnote{\url{https://hevc.hhi.fraunhofer.de/svn/svn_HEVCSoftware/tags/HM-16.20}} with the default configuration of the Low-Delay P (LDP) mode (\textit{encoder\_lowdelay\_P\_main.cfg}) at QP = 37. In this configuration, the frame at each position in every Group of Pictures (GOPs) are compressed at a fixed QP, and there is no rate control during the compression. As such, the quality fluctuates regularly in each GOP and the frame-rate has no effect for compression. This is the widely used configuration in the existing video enhancement literature~\cite{yang2017decoder, yang2018enhancing, yang2018multi, yang2019quality, guan2019mfqe, Xu_2019_ICCV}. 

\subsection{Track 3: Fixed bit-rate}

This track aims at a more practical scenario. Since HM 16.20 lacks practicability due to its high time complexity, we use the x265 library of FFmpeg\footnote{\url{https://johnvansickle.com/ffmpeg/releases/ffmpeg-release-amd64-static.tar.xz}} in this track. More importantly, rate control is a widely used strategy in video streaming to fit the compressed video to the limited band-width. Therefore, we compress the groundtruth videos by x265 with rate control at the fixed bit-rate of 200 kbps, using the following commands:

\

\noindent\texttt{ffmpeg -pix\_fmt yuv420p -s WxH -r FR -i name.yuv -c:v libx265 -b:v 200k -x265-params pass=1:log-level=error -f null /dev/null}

\

\noindent\texttt{ffmpeg -pix\_fmt yuv420p -s WxH -r FR -i name.yuv -c:v libx265 -b:v 200k -x265-params pass=2:log-level=error name.mkv}

\

\noindent In above commands, W and H are the width and height of the video, respectively. FR denotes the frame rate. Note that, we use the two-pass strategy to ensure the accuracy of rate control.

\section{Video enhancement methods}

In this section, we briefly introduce the top methods in the NTIRE 2021 challenge~\cite{yang2021ntire} and the existing methods in previous literature. The results of the challenge and the study on the proposed LDV dataset are to be detailed in Section~\ref{results}.

\subsection{Methods in the NTIRE 2021 challenge}

\begin{table*}[!t]
\footnotesize
  \centering
  \caption{The platforms, test strategies and training data of the challenge methods~\cite{yang2021ntire}}
    \begin{tabular}{ccccc}
    \toprule
    Team & Platform & GPU & Ensemble / Fusion & Extra training data\\
    \midrule
    BILIBILI AI \& FDU & PyTorch & Tesla V100/RTX 3090 & Flip/Rotation  x8 & Bilibili~\cite{bili}, YouTube~\cite{youtube} \\
    NTU-SLab & PyTorch & Tesla V100 & Flip/Rotation  x8 & Pre-trained on REDS~\cite{nah2019ntire}\\
    VUE   & PyTorch & Tesla V100 & Flip/Rotation  x8 & Vimeo90K~\cite{xue2019video} \\
    NOAHTCV  & TensorFlow & Tesla V100 & Flip/Rotation  x8 & DIV8K~\cite{gu2019div8k} (Track 2)\\
    MT.MaxClear  & PyTorch & Tesla V100 & Flip/Rotation/Multi-model x12 & Private dataset\\
    Shannon  & PyTorch & Tesla T4 & Flip/Rotation  x8 (Track 1) & -\\
    Block2Rock Noah-Hisilicon  & PyTorch & Tesla V100 & Flip/Rotation  x8 & YouTube~\cite{youtube}\\
    Gogoing  & PyTorch & Tesla V100 & Flip/Rotation x4 & REDS~\cite{nah2019ntire}\\
    NJU-Vision  & PyTorch & Titan RTX & Flip/Rotation x8 & SJ4K~\cite{song2013sjtu} \\
    BOE-IOT-AIBD  & PyTorch & GTX 1080 & Overlapping patches & - \\
    (\textit{anonymous})  & PyTorch & Tesla V100 & - & Partly finetuned from \cite{wang2019edvr} \\
    VIP\&DJI & PyTorch & GTX 1080/2080 Ti & Flip/Rotation  x8 & SkyPixel~\cite{SKYPIXEL}.\\
    BLUEDOT  & PyTorch & RTX 3090 & - & Dataset of MFQE 2.0~\cite{guan2019mfqe}\\
    HNU\_CVers  & PyTorch & RTX 3090 & Overlapping patches & - \\
    McEhance & PyTorch & GTX 1080 Ti & - & -\\
    Ivp-tencent  & PyTorch & GTX 2080 Ti & - & -\\
    \midrule
    \midrule
    MFQE~\cite{yang2018multi} & TensorFlow & TITAN Xp & - & -\\
    QECNN~\cite{yang2018enhancing} & TensorFlow & TITAN Xp & - & -\\
    DnCNN~\cite{zhang2017beyond} & TensorFlow & TITAN Xp & - & -\\
    ARCNN~\cite{dong2015compression} & TensorFlow & TITAN Xp & - & -\\
    \bottomrule
    \end{tabular}%
  \label{tab:time}%
\end{table*}%

\textbf{BILIBILI AI \& FDU Team}~\cite{yang2021ntire, bilibili} is the winner of Tracks 1 and 2 and ranks second in Track 3. They propose the Spatiotemporal Model with Gated Fusion (SMGF) for enhancing quality of compressed video. The proposed SMGF method explores the temporal correlation by selecting eight reference frames to benefit the enhancement of each target frame. The selection of reference frames considers temporal distance and QP values. The nine frames are fed into the Spaito-Temporal Deformable Fusion~(STDF)~\cite{deng2020spatio} module to explore both spacial and temporal features, and then these features are fed to the Quality Enhancement (QE) module. Additionally, they propose training two models on two datasets, and both models consist of STDF and QE. Finally, they fuse the two enhanced frames to improve the quality performance.

\textbf{NTU-SLab Team} is another winner of Track 1 and also the winner of Track 3, and ranks second in Track 2. The method proposed by the NTU-SLab Team is called BasicVSR++~\cite{yang2021ntire}, which is based on BasicVSR~\cite{chan2021basicvsr} but additionally consists of deliberate improvements on propagation and alignment. Specifically, in BasicVSR++, residual blocks are first applied to extract features from each frame. The features are then propagated under the proposed second-order grid propagation scheme, where alignment is performed by the proposed flow-guided deformable alignment. After propagation, the aggregated features are used to generate the output image through convolution and pixel-shuffling.

\textbf{VUE Team}~\cite{yang2021ntire} ranks third in Track 1. They propose combining and stacking several BasicVSR~\cite{chan2021basicvsr} models for enhancing the fidelity of compressed videos. For the perpetual track, they propose the method called ``Adaptive Spatial-Temporal  Fusion  of  Two-Stage Multi-Objective Networks''~\cite{li2021VUE}. Specifically, the first stage aims at obtaining the relatively good intermediate results with high fidelity. In this stage, a BasicVSR model is trained with the Charbonnier loss~\cite{charbonnier1994two}. At the second stage, they train two BasicVSR models for different refinement purposes. One refined BasicVSR model is trained with a combination of the Charbonnier loss~\cite{chan2021basicvsr} and LIPIS loss~\cite{zhang2018unreasonable}.
Another refined BasicVSR model is only trained with the LPIPS loss~\cite{zhang2018unreasonable}. Then, to fuse the results from the two networks, they design a novel adaptive spatial-temporal fusion scheme, which is able to eliminate flickering effects.

\textbf{NOAHTCV Team}~\cite{yang2021ntire} ranks third in Track 2. They follow MFQE~\cite{yang2018multi} to use Peak Quality Frames (PQFs) to facilitate the enhancement of other frames. The target frame and two neighboring PQFs are fed into the proposed network. They first use a deep network to estimate the offsets of the three input frames, and then the offsets are used to align and fuse the features. After the initial feature extraction and alignment, a multi-head U-Net is applied to process each feature, and they fuse the features with scale-dependant deformable convolutions at each scale. Finally, the output features are fused and fed into residual blocks to reconstruct the enhanced frame. They use the $L2$ loss for Tracks 1 and 3, and the GAN Loss + Perceptual loss + $L2$ loss for Track~2.

\textbf{MT.MaxClear Team}~\cite{yang2021ntire} ranks third in Track 3. They propose utilizing Charbonnier penalty loss, DCN offsets Total Variation loss and DCN offsets Variation loss to the EDVR~\cite{wang2019edvr} network to improve the stability of training. DCN offsets Total Variation loss encourages the predicted DCN offsets are smooth in spatial space. DCN offsets Variation loss encourages the predicted DCN offsets between different channels do not deviate too much from the offsets mean. In Track 2, they add the sharpening operation on the enhanced frames for better visual perception.

\begin{figure*}[!t]
\centering
\subfigure[PSNR on Track 1 (unprocessed video = 30.54 dB)]{\includegraphics[width=.46\linewidth]{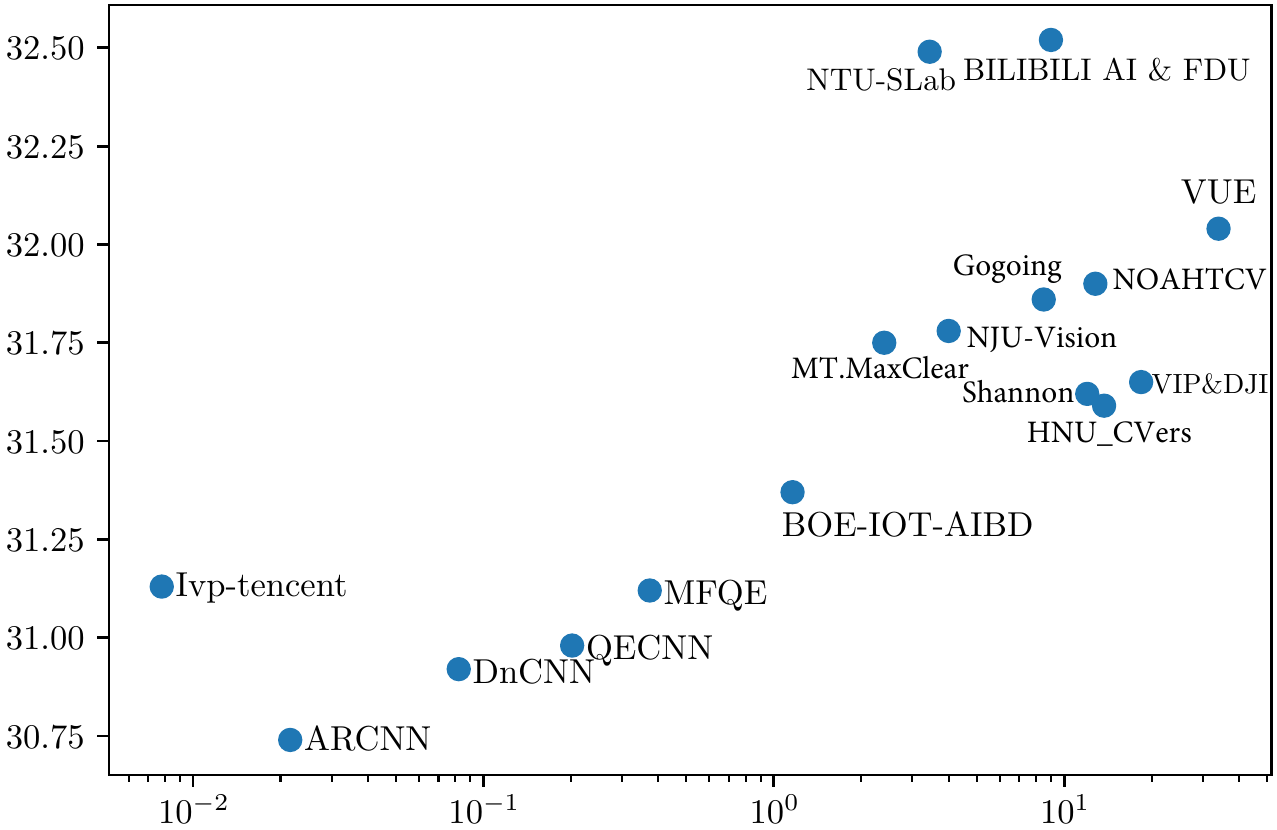}}
\hspace{1em}
\subfigure[MS-SSIM on Track 1 (unprocessed video = 0.9305)]{\includegraphics[width=.46\linewidth]{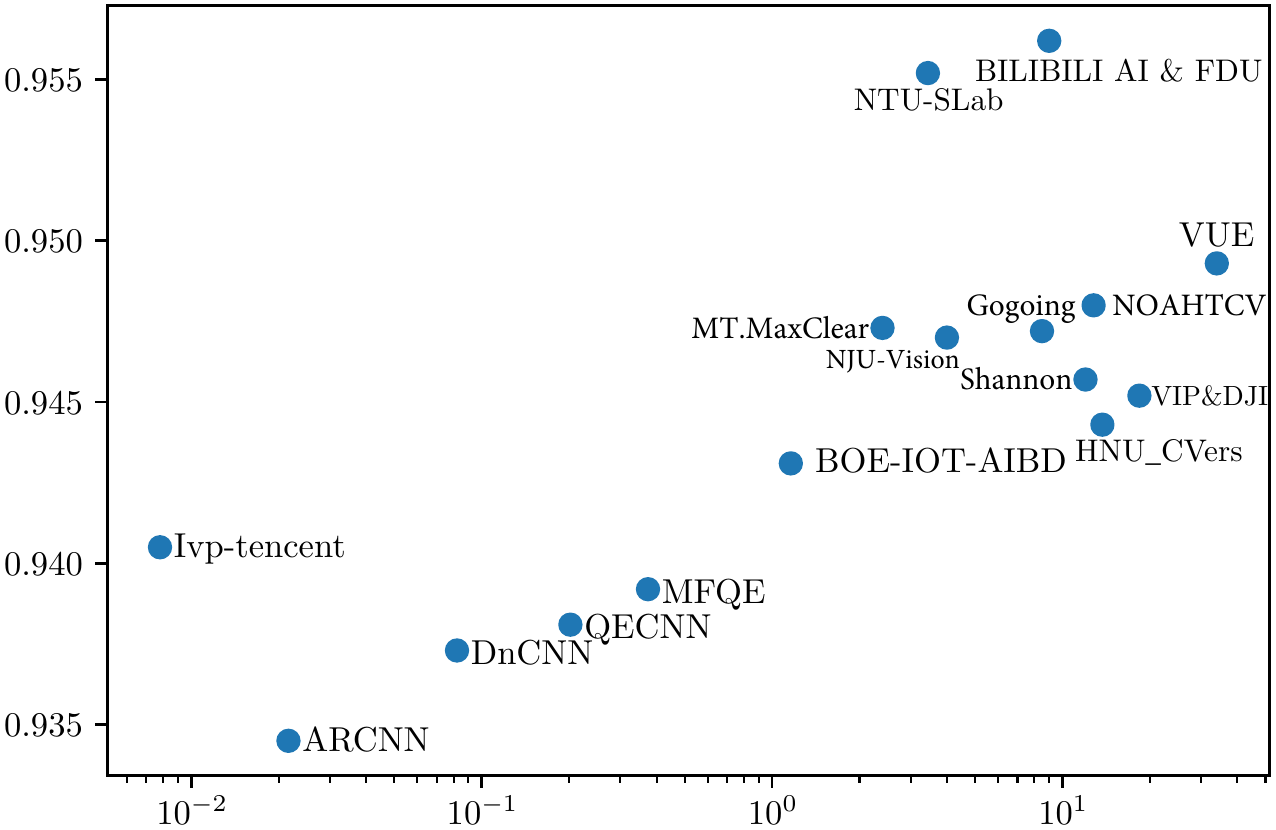}}
\caption{The results on Track 1. The running time is reported by the authors of the teams. The platforms used for test are listed in Table~\ref{tab:time}.}
\label{fig:track1}
\end{figure*}

\subsection{Methods in previous literature}

\textbf{ARCNN}~\cite{dong2015compression} and \textbf{DnCNN}~\cite{zhang2017beyond} are the deep neural networks proposed for reducing the compression artifacts of JPEG images. ARCNN is the first method in this direction, which consists of four convolutional layers. DnCNN is a deeper network with 20 convolution layers and uses batch normalization. DnCNN learns to estimate the residual image, instead of directly reconstructing the enhanced image in ARCNN. These image enhancement methods are the fundamental for video enhancement. In this report, we re-train these networks by the training videos of our LDV dataset.

\textbf{QE-CNN}~\cite{yang2018enhancing, yang2017decoder} is a single-frame quality enhancement method proposed for enhancing HEVC compressed video. In QE-CNN, a five-layer convolutional neural network, called QE-CNN-I, is designed to enhance I-frames. Moreover, additional five CNN layers are proposed to extract and handle the features of inter-coding frames, which are then combined with the intra-coding features extracted by QE-CNN-I to enhance the quality of B- and P-frames in HEVC. The network for enhancing B- and P-frames is named as QECNN-B. The original QE-CNN is designed for only enhancing the Y channel. In this report, we re-train QE-CNN on the LDV training videos in the RGB domain.

\textbf{MFQE}~\cite{yang2018multi, guan2019mfqe} is proposed to enhance compressed video taking advantage of higher quality neighboring frames. In MFQE, the Peak Quality Frames (PQFs) are defined as the frames with higher quality then its neighboring frames, and PQFs are used to benefit the enhancement of other frames. In MFQE, an SVM-based~\cite{yang2018multi} or LSTM-based~\cite{guan2019mfqe} PQF detector is first used to detect the PQFs, and then the MF-CNN, which contains a motion compensation network and a quality enhancement network, is utilized to enhance the quality of non-PQFs. The same as QE-CNN, the MFQE model is re-trained for RGB frames on the training set of LDV.


\begin{figure*}[!t]
\centering

\subfigure[MOS $\uparrow$ on Track 2]{\includegraphics[width=.465\linewidth]{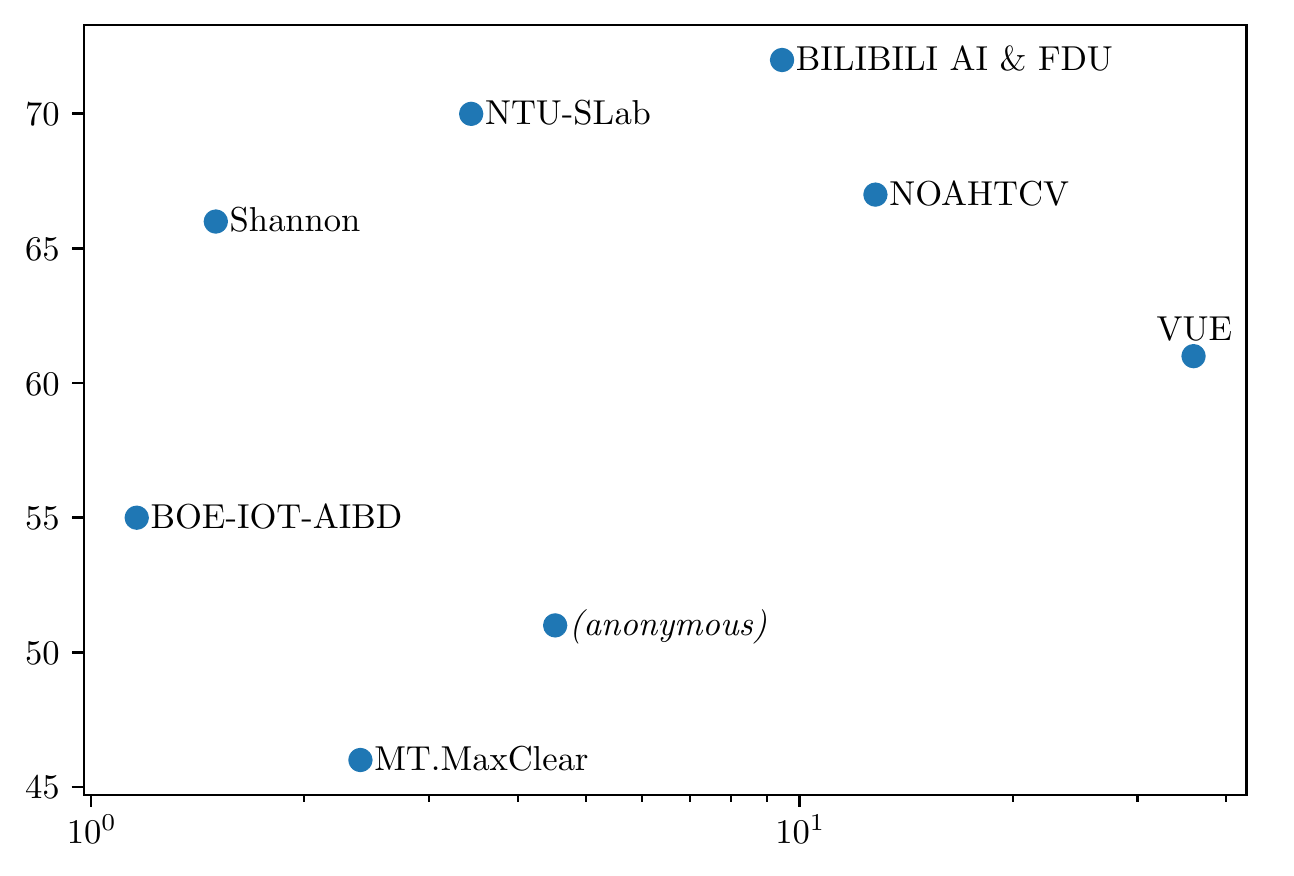}}
\hspace{1em}
\subfigure[VMAF $\uparrow$ on Track 2]{\includegraphics[width=.465\linewidth]{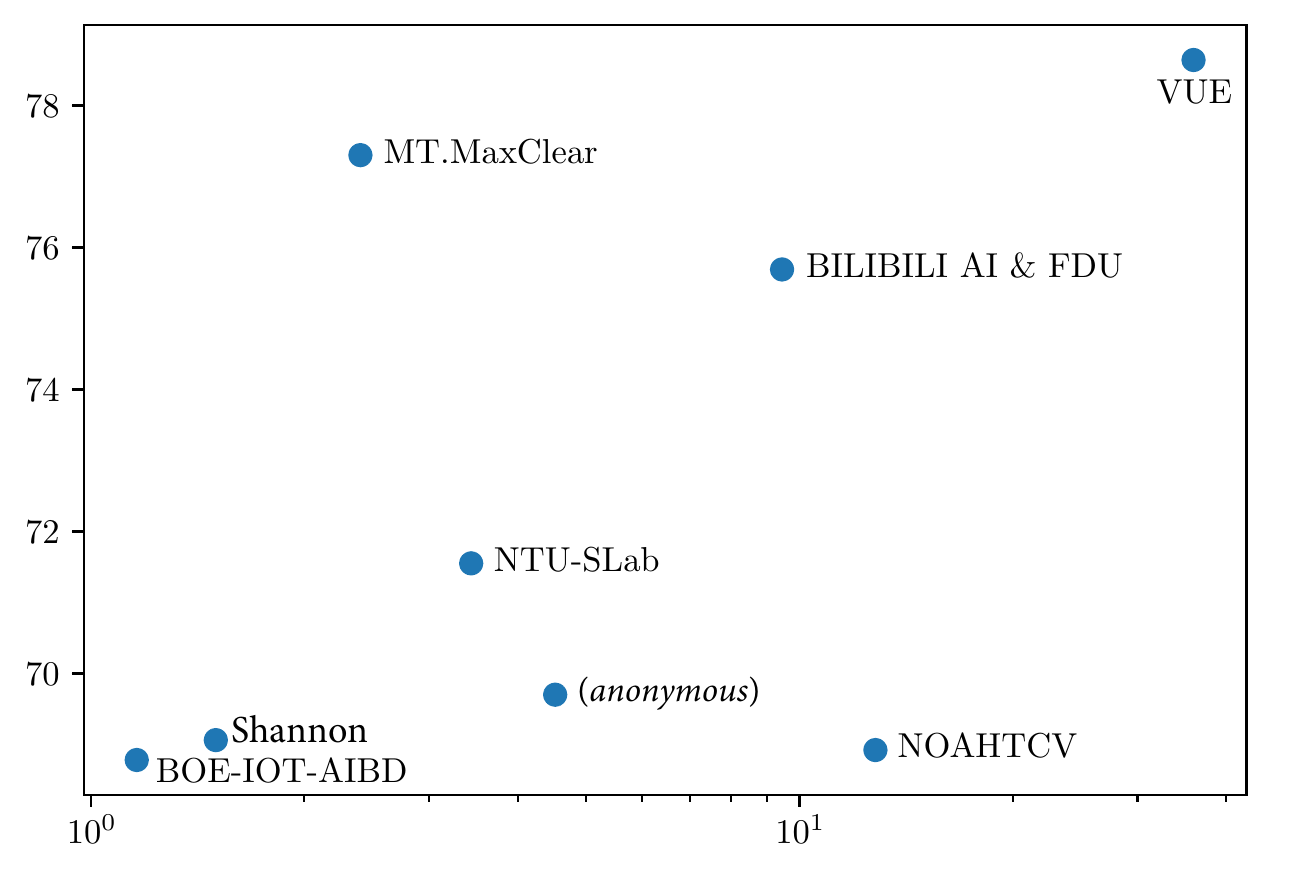}}

\subfigure[LPIPS $\downarrow$ on Track 2]{\includegraphics[width=.475\linewidth]{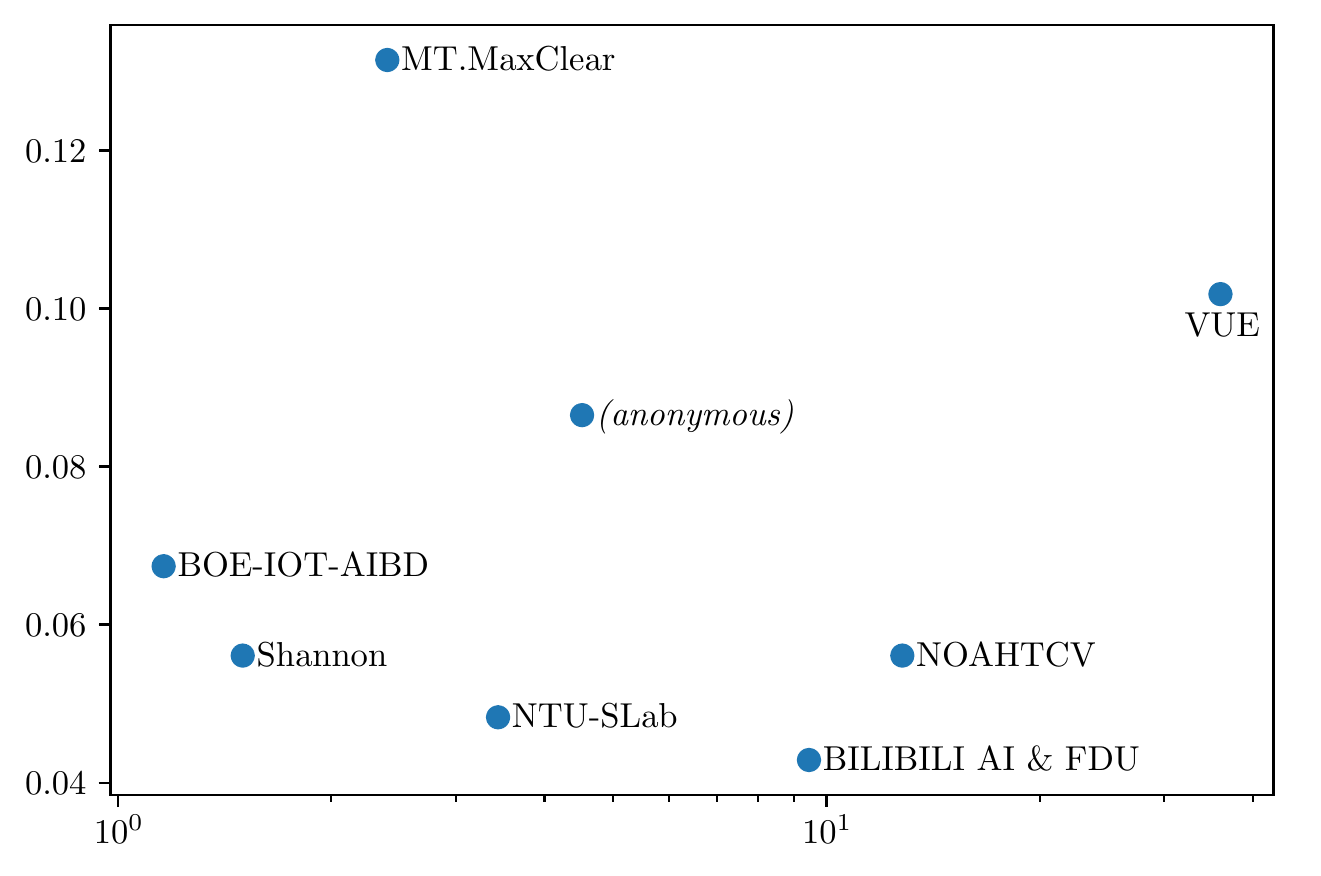}}
\hspace{1em}
\subfigure[FID $\downarrow$ on Track 2]{\includegraphics[width=.465\linewidth]{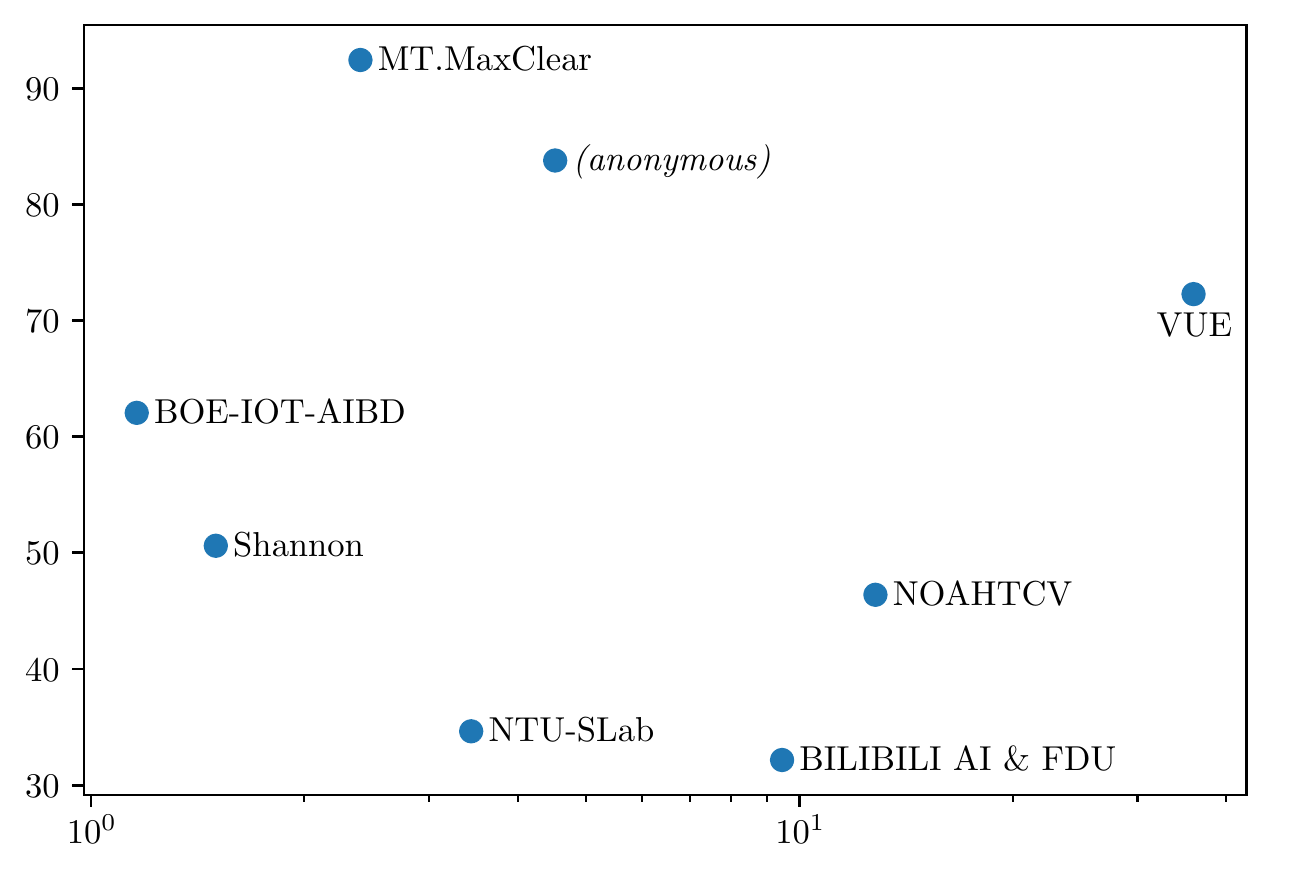}}

\caption{The results on Track 2. The running time is reported by the authors of the teams. The platforms used for test are listed in Table~\ref{tab:time}. }
\label{fig:track2}
\end{figure*}

\section{Challenge results and studies on LDV}\label{results}

\subsection{Performance on Track 1}

The quality performance vs. running time of Track 1 is illustrated in Figure~\ref{fig:track1}. It can be seen that the methods proposed in NTIRE challenge obviously outperform the previous methods, and therefore advance the state-of-the-art of quality enhancement for compressed video.

\textbf{PSNR vs. time efficiency.} In the proposed methods, the BILIBILI AI \& FDU and NTU-SLab Teams achieve the best quality with similar PSNR performance. The VUE Team ranks in the following at the third place. The NTU-SLab Team has the fastest running speed among the top 5 methods, and therefore makes a good trade-off between quality performance and time efficiency. The most time-efficient method is Ivp-tencent, which is able to enhance video frames at 120 frames per second. The speed of Ivp-tencent significantly outperforms other methods, while its quality performance is comparable with and slightly higher than MFQE~\cite{yang2018multi}.

\textbf{MS-SSIM performance.} Figure~\ref{fig:track1}-(b) shows the MS-SSIM performance on Track 1. We can observe from Figure~\ref{fig:track1}-(b) that the MS-SSIM performance is generally consistent with the PSNR performance in Figure~\ref{fig:track1}-(a). This may indicate that when optimizing the enhancement networks for PSNR, the MS-SSIM performance will be normally consistent with PSNR.

\begin{table}[!t]
\small
  \centering
  \caption{The normalized and non-normalized MOS values.}
    \begin{tabular}{ccc}
    \toprule
    Team  & MOS (norm) & MOS (w/o norm) \\
    \midrule
BILIBILI AI \& FDU Team & 72.2  & 70.0 \\
    NTU-SLab & 70.1  & 68.8 \\
    NOAHTCV & 67.3  & 66.9 \\
    Shannon & 65.8  & 65.9 \\
    VUE   & 60.7  & 62.0 \\
    BOE-IOT-AIBD & 54.5  & 57.8 \\
    \textit{(anonymous)} & 50.9  & 55.5 \\
    MT.MaxClear & 46.2  & 52.1 \\
    \midrule
    \midrule
    Unprocessed video & 36.0  & 44.5 \\
    \bottomrule
    \end{tabular}%
  \label{tab:MOS}%
\end{table}%

\subsection{Performance on Track 2}

The quality performance vs. running time of Track 2 is shown in Figure~\ref{fig:track2}. Note that in the four perceptual metrics, the better MOS and VMAF values indicate better quality, while lower LPIPS and FID indicate better quality. 

\textbf{Normalized vs. non-normalized MOS.} As mentioned in Section~\ref{VQA}, we normalize the MOS values of each subject by \eqref{norm} and rank the methods according to the average normalized MOS. We compare the results of normalized MOS and non-normalized MOS in Table~\ref{tab:MOS}. It can be seen from Table~\ref{tab:MOS} that the normalization on MOS does not change the ranking of the proposed methods, but only enlarges the interval between the lowest and highest scores and therefore makes the difference between the methods more clear. Therefore, in the NTIRE 2021 challenge, we use the normalized MOS for Track 2. The results are shown in  Figure~\ref{fig:track2}.

\begin{figure*}[!t]
\centering
\subfigure[PSNR on Track 3 (unprocessed video = 28.34 dB)]{\includegraphics[width=.453\linewidth]{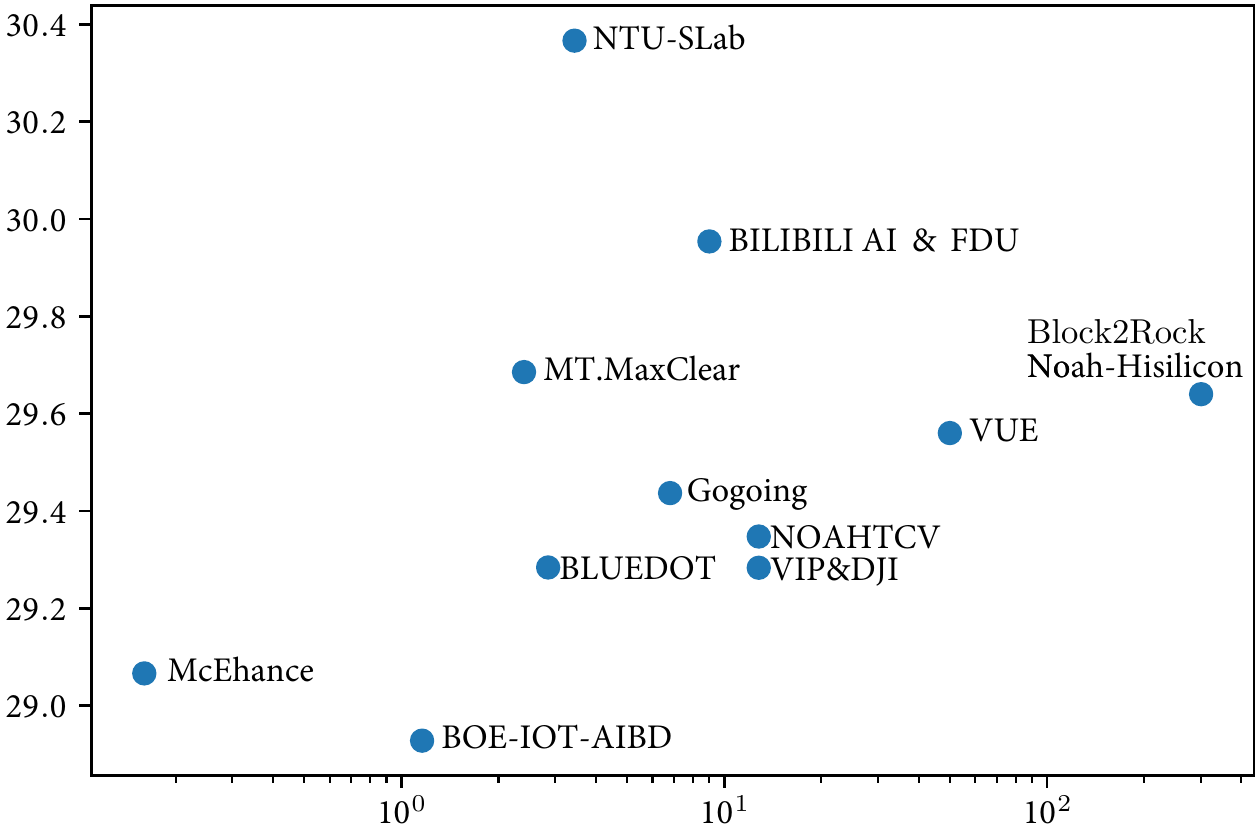}}
\hspace{1em}
\subfigure[MS-SSIM on Track 3 (unprocessed video = 0.9243)]{\includegraphics[width=.46\linewidth]{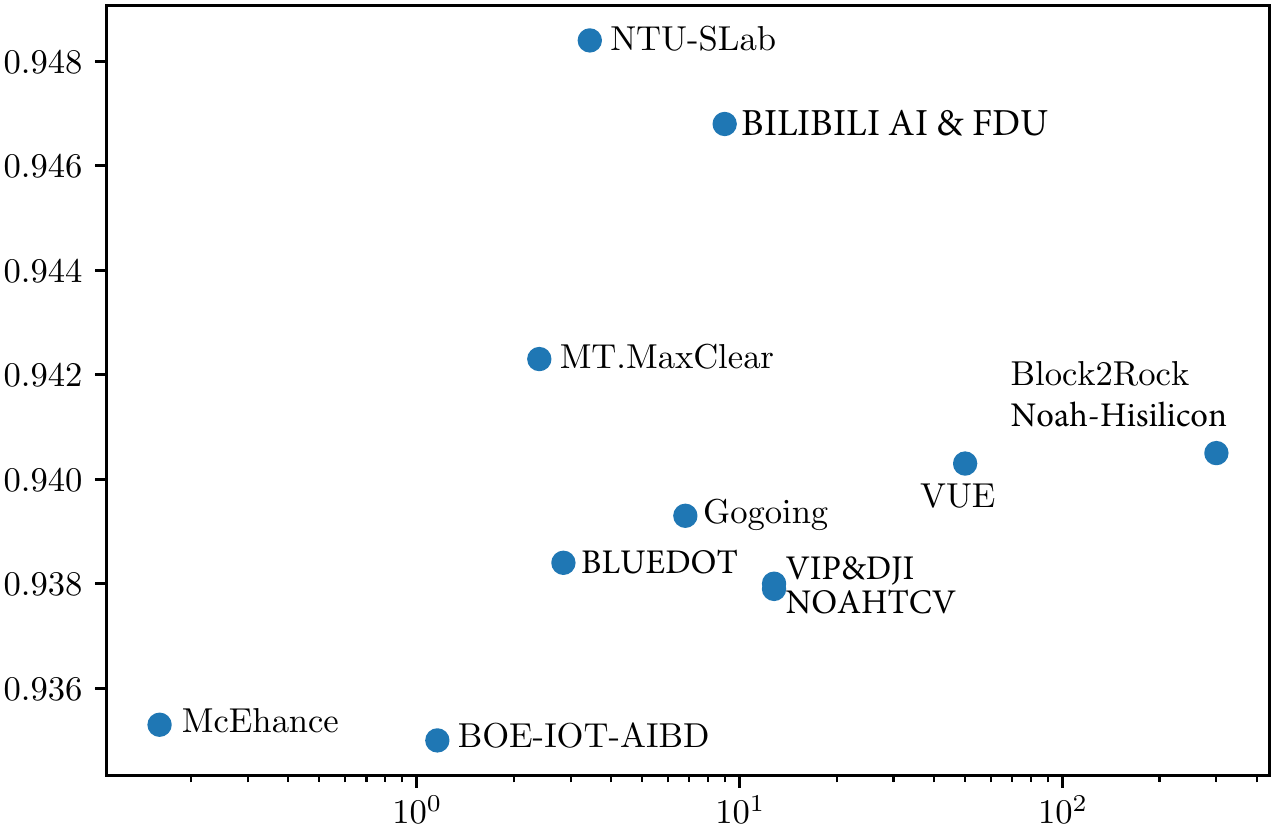}}
\caption{The results on Track 3. The running time is reported by the authors of the teams. The platforms used for test are listed in Table~\ref{tab:time}.}
\label{fig:track3}
\end{figure*}

\textbf{MOS vs. time efficiency.} As shown in Figure~\ref{fig:track2}-(a), BILIBILI AI \& FDU, NTU-SLab and NOAHTCV Teams achieve the top perceptual quality in terms of MOS. BILIBILI AI \& FDU ranks at the first place in this track. The NTU-SLab and NOAHTCV Teams rank second and third, respectively. The running time of NTU-SLab is several times lower than BILIBILI AI \& FDU and NOAHTCV. Therefore, NTU-SLab achieves a good balance between perceptual quality and time efficiency. The Shannon Team has faster speed than NTU-SLab but with lower MOS value. Moreover, it can be seen from Table~\ref{tab:time} that all top methods use the ensemble strategies~\cite{timofte2016seven} during inference and enlarge the training set by adding extra videos for training. These may indicate the effectiveness of the ensemble strategy for perceptual enhancement and show the benefit of large-scale training data.

\textbf{Other metrics. } As Figures~\ref{fig:track2}-(c) and -(d) show, the rankings of the top four methods in terms of LPIPS and FID are the same as MOS. This shows that the perceptual metrics are effective for evaluating subjective quality. However, the LPIPS and FID values of unprocessed video are 0.0752 and 48.94, respectively. Therefore, some methods even have worse LPIPS and FID performance than the unprocessed videos, but their MOS values are all obviously better the unprocessed videos (MOS = 36). Also, the ranking on VMAF is different from that on MOS. This may also indicate the limited reliability of the perceptual metrics, and the metrics LPIPS and FID which are designed for image may be not very suitable for evaluating video.  

\subsection{Performance on Track 3}

The PSNR and MS-SSIM performances on Track 3 are illustrated in Figure~\ref{fig:track3}. The NTU-SLab, BILIBILI AI \& FDU and MT.MaxClear Teams rank at the first, second and third places in terms of both PSNR and MS-SSIM. They improve the PSNR of unprocessed video (28.34 dB) by 2.03 dB, 1.62 dB and 1.35 dB, respectively.
In these teams, the first-ranked team NTU-SLab also achieves good time efficiency. MT.MaxClear is the fastest among the top 3 methods, while the McEnhance Team is the most time-efficient in all methods. It can be seen from Figure~\ref{fig:track3} that the ranks of PSNR and MS-SSIM are consistent. The same as Track 1, this may show that the MS-SSIM performance will be normally consistent with PSNR, in the case that the quality enhancement networks are optimized towards PSNR.

\section{Analysis on training data}

\begin{figure}[!t]
\centering
\includegraphics[width=1\linewidth]{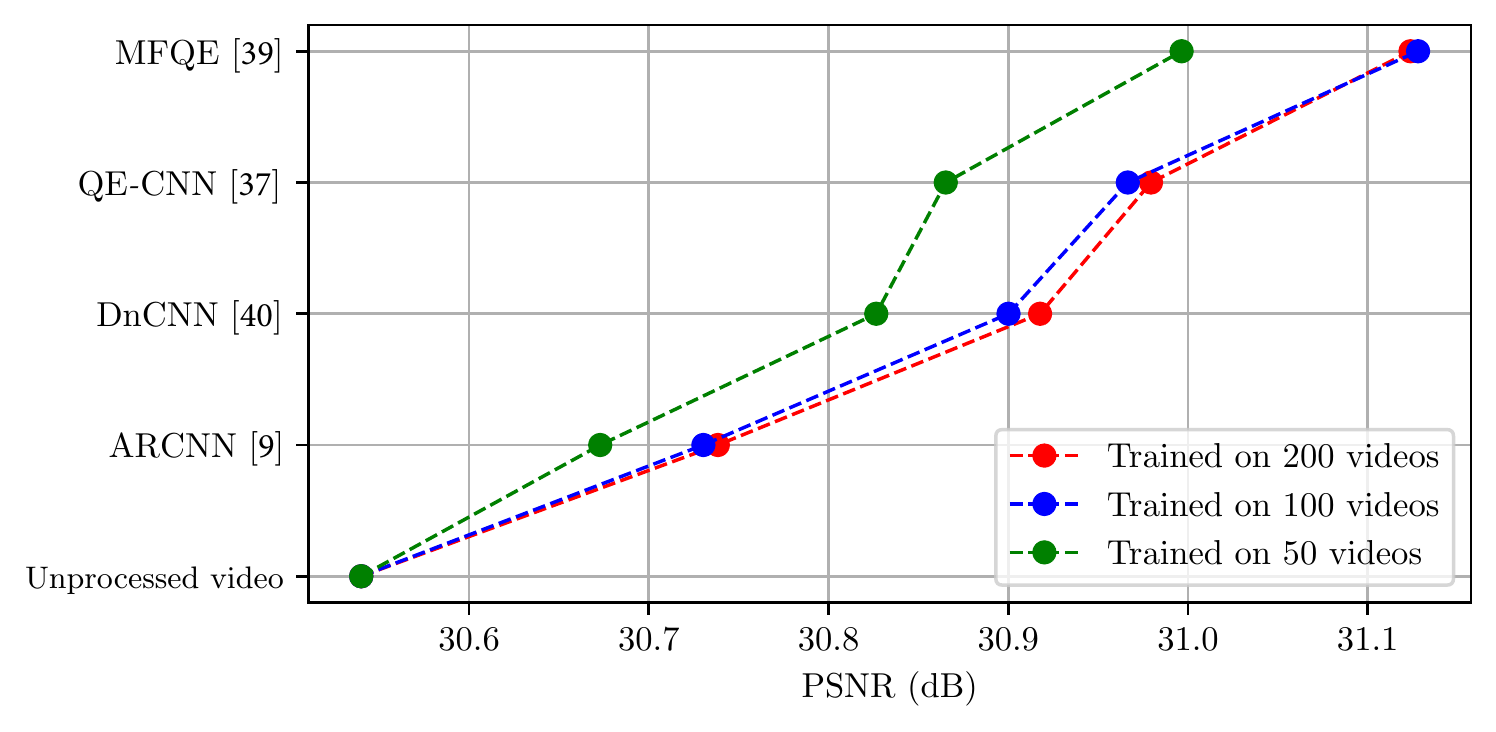}
\caption{The PSNR performance on the test set of the models trained by different scales of data.}
\label{fig:train}
\end{figure}

Finally, we analyze the scale of training data and the impact on the test performance. Figure~\ref{fig:train} illustrates the PSNR results on the test set for different scales of training videos. The red line indicates the performance when the models are trained on all 200 training videos of the proposed LDV dataset. The blue and green lines show the results when training on 100 and 50 randomly selected videos from the training set of LDV, respectively. It can be seen from Figure~\ref{fig:train} that enlarging the scale of training data
from 50 to 100 obviously improves the test results, and further enlarging the number of training video to 200 also further slightly advances the performance.

As mentioned in Section~\ref{intro}, the training datasets used in previous methods contain less than 150 videos. Therefore, the proposed LDV dataset may have the potential to advance the performance of the existing methods. However, in Figure~\ref{fig:train}, we also notice that the performance of MFQE~\cite{yang2018multi} does not improve when increasing the number of training videos from 100 to 200. This may be because of the bias in the data, and it indicates that the scale of the proposed LDV dataset is still incremental, especially the test set. It is an interesting future work to further extend the LDV dataset and establish benchmarks on datasets with larger scale.

\section{Conclusion}

This paper has introduced the LDV database, which is a new large-scale dataset with diverse videos. The proposed LDV dataset contains 240 videos with diverse categories of scenes, different kinds of motions and various frame-rates. The LDV dataset has been used in the video enhancement challenge of NTIRE 2021. We analyzed the proposed methods in the challenge and the methods in previous literature on the LDV dataset. These methods and results established a benchmark for quality enhancement of compressed video. We hope that the proposed LDV dataset and the benchmark of this challenge are able to provide solid baselines for the quality enhancement of compressed video and further benefit the future research in this field.

\section*{Acknowledgments}

We thank the NTIRE 2021 sponsors: Huawei,
Facebook Reality Labs, Wright Brothers Institute, MediaTek, OPPO and ETH Zurich (Computer Vision Lab). We also thank the volunteers for the perceptual experiment of Track 2.

{\small
\bibliographystyle{ieee_fullname}
\bibliography{egbib}
}

\end{document}